\documentclass[fleqn,usenatbib,onecolumn]{mnras}

\usepackage{epsfig,amsmath,subfigure}
\usepackage{blindtext}
\usepackage[ugly]{nicefrac}
\usepackage{multirow,makecell}

\def\Aa{{\it Astron. Astrophys.} \,}

\def\apj{{\it ApJ \,}}

\def\apjl{{\it Ap. J. Lett.} \,}

\def\prd{{\it Phys. Rev. D.} \,}
\def\mn{{\it MNRAS} \,}

\def\rmph{{\it Rev. Mod. Phys.} \,}

\def\ea{\ et al. \,}

\def\be{\begin{equation}}
\def\ee{\end{equation}}

\def\g{\gamma}
\def\gt{\gamma_t}

\def\pe{proton and electron }
\def\pes{protons and electrons }
\def\spp{\sigma_{pp}^{\pi^0,\pi^{\pm}}}

\def\gp{\gamma_p}
\def\rd{r_d}

\def\zo{t_o}

\def\sinel{\sigma_{pp}^{in}}

\DeclareMathOperator{\erf}{erf} 
\DeclareMathOperator{\erfi}{erfi}

\title[Energetic Particles in Halos of SFGs]{Energetic Particles in Halos of 
Star Forming Galaxies}
\author{Yoel Rephaeli}

\author[Yoel Rephaeli, Sharon Sadeh]{Yoel Rephaeli$^{1,2}$\thanks{E-mail:
yoelr@tauex.tau.ac.il}, Sharon Sadeh$^{1}$\\
$^{1}$School of Physics and Astronomy, Tel Aviv University, Tel Aviv, 69978, Israel\\
$^{2}$Center for Astrophysics and Space Sciences, University of California,
San Diego, La Jolla, CA 92093-0424}

\date{Accepted XXX. Received YYY; in original form ZZZ}

\pubyear{2019}

\begin{document}

\pagerange{\pageref{firstpage}--\pageref{lastpage}}
\maketitle

\label{firstpage}

\begin{abstract}
Quantitative modeling of the spectro-spatial distributions of energetic electrons and protons 
in galactic halos is needed in order to determine their interactions with the local plasma and 
radiation fields, and also to estimate their residual spectral densities in intracluster and 
intergalactic environments. We develop a semi-analytic approach for calculating the particle distributions in the halo based on a detailed diffusion model for particle propagation from 
acceleration sites and interactions in the galactic disk. Important overall normalization of 
our models is based on results from detailed modeling in the Galactic disk with the GALPROP 
code. This provides the essential input for determining particle distributions in the outer disk, 
which are used as source terms for calculating the distributions in the extensive halo for a range 
of values of key parameters affecting energy losses and propagation mode. Our modeling 
approach is applied to the two edge-on star-forming galaxies NGC 4631 and NGC 4666, for 
which recent mapping of radio emission in the inner halo provides the required overall 
normalization. We predict the levels and spatial profiles of radio, X-ray, and $\gamma$-ray 
emission in the halos of these galaxies. Our quantitative modeling enables us to estimate the 
total calorimetric efficiencies of electrons and protons in star-forming galaxies, and to predict 
their residual spectral distributions in the outer halo and intergalactic space.
\end{abstract}



\section{Introduction}

Measurements of energetic particles (`cosmic rays'), their radiative yields, and of magnetic 
fields in various media in interstellar space of the Galaxy have led to reasonably good 
knowledge of the particle spectro-spatial distributions (SSDs) in the Galactic disk, where 
the stellar particle acceleration sources are located. Particle propagation modes and energy 
loss processes are sufficiently well understood to expect that energetic electrons lose a large 
fraction of their energy by the time they reach the outer disk. This fraction is much smaller 
for protons whose (effective) energy loss timescale is much longer than a typical escape time 
from the disk. The distributions of both particles in the Galactic halo are of appreciable 
interest as they provide an essential basis for estimating particle escape rates from 
star-forming (SF) and starburst (SB) galaxies, and for quantifying their radiative yields in 
galactic halos (and IG space). 

Outside the galactic disk, the couplings of protons and electrons to the lower density gas, 
weaker magnetic fields, and less intense radiation fields, are reduced but are expected to 
be still significantly stronger than in IG space. There has been little quantitative modeling 
of energetic particles in galactic halos, in spite of the fact that determining the particle 
distributions in these extended regions is needed for estimating their contributions to 
the mean IG spectral density. Additionally, the radiative yields in halos have to be estimated 
for separating these out from their respective values in the disk to enable precise determination 
of the latter, quite likely dominant emission. With increasing knowledge of the astrophysical environment in galactic halos, there is sufficient motivation for a more complete assessment 
of galactic energetic particles and their full emission in the disk and halo. 

The radiative yields of electrons (by bremsstrahlung, Compton, and synchrotron processes) 
and protons, whose interactions with ambient protons produce neutral ($\pi^{0}$) and 
charged ($\pi^{\pm}$) pions which decay into $\gamma$-ray photons, electrons and positrons, 
respectively, cover the wide spectral range from radio to $\gamma$-ray. Measurement of this 
emission can potentially provide sufficient quantitative basis for determining the particle SSDs 
in the Galactic halo, in the halos of other SFGs, and in SBGs.  

Modeling \pe spectral and spatial properties in the halo requires detailed knowledge 
of their SSDs in the disk. While semi-analytic calculations of these distributions have 
been made, mostly based on diffusion models (e.g., Syrovatskii 1959), currently the most 
realistic description of particle propagation in the Galactic disk is provided by the 
GALPROP code with which several  diffusion (with and without re-acceleration) models 
were explored (Strong \ea 2010). For various purposes, it is useful to have a semi-analytic 
alternative to this code that can be easily employed to study predictions of various models 
for the spatial profiles of the gas density and magnetic field in the outer disk and halo, 
where the magnetized fully ionized gas properties and particle propagation modes are 
considerably simpler to model than in the inner much more complex disk.

A complete description of energetic particles in disk and inner halo of a SF or SB 
galaxy will provide the basis for determining their distributions in the outer halo 
and for estimating their contributions to the overall particle energy densities in IG 
space. For this purpose, an efficient, easily adjustable semi-analytic calculation 
scheme of energetic particle contributions by SF galaxies of various classes is of 
much interest in the context of a statistical study of the more general open issue of 
the origin and IG propagation of energetic particles. Results of such a study will 
also have important ramifications for understanding the origin of energetic electrons 
whose large-scale radio (`halo') emission has been mapped in many galaxy clusters. 
For example, the number of SFGs in a rich cluster may be sufficiently large for 
these to constitute an appreciable fraction of the emitting electrons in radio `halos' 
(Rephaeli \& Sadeh 2016).

The motivation for detailed modeling of \pe SSDs in the outer disk and halo of a SF 
galaxy has recently been strengthened by radio measurements of the emission in the 
outer disks and inner halos of a sample of 35 nearby edge-on SFGs by the EVLA 
survey CHANG-ES  (Wiegert \ea 2015). Analysis of these measurements yielded the 
first co-added median map of the inner halo emission of a spiral galaxy at 1.5 GHz. 
This map of the emission profile well beyond the galactic disk provides a quantitative 
basis for a reliable normalization of the SSD of primary and secondary electrons in the 
galactic halo, and for a realistic assessment of the predicted emission in the X-ray 
and $\gamma$-ray regions.

Here we describe \pe SSDs in the halo of a SFG based on a diffusion model for their 
propagation in the disk and halo as they traverse the magnetized gas and lose energy by 
all the relevant processes. In Section 2 we present the basic model and results of its 
implementation to the two nearby SFGs NGC 4631 and NGC 4666 for which radio 
emission has been measured in the inner halo. Our quantitative treatment provides a 
reliable basis for predicting the levels of halo radio, X-\&-$\gamma$-ray emission in 
these galaxies, as detailed in Section 3, and further discussed in Section 4.

\section{Particle Distributions and Radiative Yields}

Particle acceleration in SFGs is largely a stellar-related phenomenon driven by shocks 
in SN remnants and pulsar wind nebulae, with the sources distributed across the stellar 
disk. The particles lose energy by interactions with the magnetized IS gas and radiation 
fields as they diffuse out of the acceleration sites across the disk, eventually reaching the 
outer disk and thereafter continue their (faster) random walk in the dilute halo. Extensive 
semi-analytic and numerical modeling of the particle propagation, energy losses, and their 
steady-state SSDs has focused on the complex disk environment where consequences of 
their interactions with IS media have either been measured (e.g., radio emission) or 
indirectly deduced (e.g., gas ionization and heating). Contrasting detailed models for the 
particle radiative yields with observations provides the basis for determining the particle 
distributions across the entire disk.

Current knowledge of particle distributions in galactic halos is poor, limited as it is to 
rough estimates based on qualitative considerations (based mainly on estimates of particle 
`residence times' in the disk and `calorimetric' fractions). A quantitatively reliable 
determination of the particle SSDs in the outer disk provides the essential input needed in 
the calculation of the distributions in the halo, and for (e.g.) estimating the residual particle 
spectral flux in the outer halo. Extension of the quantitative treatment of particle distributions 
to the halo by using detailed results of the distributions in the outer disk is a basic objective 
of our work.

Clearly, a detailed treatment of the inherently complex IS environment traversed by 
energetic particles, and the wide ranges of all the relevant quantities that affect their 
energy losses and propagation modes, necessitate an extensive numerical code to 
determine the particle SSDs. This provided the main motivation for developing the 
GALPROP code (e.g., Strong \ea 2010)\footnote{The current version of GALPROP 
is available at https://gitlab.mpcdf.mpg.de/aws/galprop}, currently the most 
comprehensive program which is widely used in modeling non-thermal (NT) 
phenomena in the Galaxy. The code is based on a solution to the kinetic equation 
for the particle steady-state distributions, asymptotically attained following diffusion 
of the particles from their acceleration sites into the full disk while losing energy by 
the various processes (and possibly also gaining energy by re-acceleration). 

Because our main objective is a reasonably realistic modeling of \pe distributions 
and their predicted radiative yields in the galactic halo, a more optimal approach 
for our purposes is a simpler semi-analytic treatment that can be easily implemented 
in comparison with the use of a more detailed (but cumbersome) full-fledged code. 
However, we do use detailed results from extensive GALPROP modeling of the 
Galaxy in order to normalize our models to ensure consistency with observational 
data. The most relevant GALPROP-based study is that of Strong \ea (2010), who 
explored several diffusion models for energetic \pe spectral distributions in the 
Galactic disk, and determined their bolometric radio, X-$\&-\gamma$-ray yields. 
(For more recent GALPROP-based studies of the Galaxy, see Orlando \& Strong 2013, 
and Orlando 2019.)  As we describe below, we consider similar models to those 
explored by Strong \ea (2010), and adopt the same initial \pe power-law spectra and 
source distribution (which is based on the observed pulsar population).

The above modeling approach is applied next to the two SFGs NGC 4631 and NGC 4666, 
for which there are no observationally-normalized GALPROP models for comparison. 
However, the radio intensity maps for the disk and inner halo (Wiegert \ea 2015) -- the 
main motivation for our selection of these nearby edge-on galaxies for this study -- provide 
the basic normalization of our models. Other available observationally determined model 
parameters for each galaxy are the SF rate, mean magnetic field in the disk, and luminosities 
in the visible and FIR bands.

\subsection{Particle Distributions}

The spatial distribution of particle sources (e.g., SNR, pulsar wind nebulae) can be 
modeled by a Gaussian (e.g., Syrovatskii 1959) which, for an ellipsoidal disk morphology 
with semi-major and semi-minor axes $a$ and $b$, is 
\begin{equation}
Q(\vec{r_i})=\frac{1}{\pi^{3/2}}\frac{1}{a^2b}
e^{-\nicefrac{x_i^2+y_i^2}{a^2}-\nicefrac{z_i^2}{b^2}},
\label{eq:cr_src}
\end{equation}
where $\vec{r_i}\equiv(x_i,y_i,z_i)$ is the position vector of source $i$ in Cartesian 
coordinates.

The SSD of primary electrons at the post-diffusion position vector $\vec{r_o}$ and 
time $t_o$ is the Green function solution of the diffusion equation (Atoyan, Aharonian, 
\& V\"{o}lker 1995):
\begin{equation}
f_e(\vec{r_o},\zo,\gamma)=\int_0^{t_o}\frac{\Delta \Dot{N_e}(\gt)P_e(\gt)}
{\pi^{3/2}P_e(\gamma)\rd^3}e^{-\nicefrac{r^2}{\rd^2}}dt_i,
\label{eq:gr_fnc}
\end{equation}
where $r\equiv|\vec{r_o}|$, $\gamma$ and $\gamma_t$ are, respectively, the post-and pre-diffusion energies of the electron, $\Dot{N_e}(\gt)$ is the spectrum of the source-injected electrons, P$_e$($\gamma$) is the energy loss rate, and $r_d$ is the diffusion radius. 
Combining equations~(\ref{eq:cr_src}) and~(\ref{eq:gr_fnc}) as
\begin{eqnarray}
f_e(\vec{r_o},t_o,\gamma)&=&\frac{1}{\pi^{3/2}}\frac{1}{a^2b}\int_0^{t_o}dt_i\frac{\Delta \Dot{N_e}(\gt)P_e(\gt)}{\pi^{3/2}P_e(\gamma)\rd^3}\int\int\int_V
e^{-\nicefrac{(x_o-x_i)^2+(y_o-y_i)^2}{r_d^2}-\nicefrac{(z_o-z_i)^2}{r_d^2}}
e^{-\nicefrac{x_i^2+y_i^2}{a^2}-\nicefrac{z_i^2}{b^2}}dx_idy_idz_i,
\end{eqnarray}
yields the total post-diffusion distribution of primary electrons at $\vec{r_o}$ and 
time t$_o$. With the steeply varying Gaussian source distribution, the volume-integration 
limits can be changed to infinity without appreciable loss of accuracy; doing so results in 
\begin{eqnarray}
f_e(\vec{r_o},t_o,\gamma)&=&\frac{1}{\pi^{3/2}}
\frac{\Delta \Dot{N_e}(\gt)P_e(\gt)}{P_e(\g)(a^2+r_d^2)\sqrt{b^2+r_d^2}}
e^{-\nicefrac{x_o^2+y_o^2}{a^2+r_d^2}-\nicefrac{z_o^2}{b^2+r_d^2}}.
\label{eq:primel}
\end{eqnarray}

The starting point for calculation of the secondary electron SSD is the equivalent solution 
for source-injected protons, which describes the distribution of post-diffusion protons just 
before they interact with protons in the ambient gas. Secondary electrons and positrons are 
produced in these proton-proton interactions via muon decays (following their production 
in charged pion decays), at position $\vec{r_m}$ and time t$_m$
\begin{eqnarray}
f_p(\vec{r_m},t_m,\gamma)&=&\frac{1}{\pi^{3/2}}
\frac{\Delta \Dot{N_p}(\gt)P_p(\gt)}{P_p(\g)(a^2+r_d^2)\sqrt{b^2+r_d^2}}
e^{-\nicefrac{x_m^2+y_m^2}{a^2+r_d^2}-\nicefrac{z_m^2}{b^2+r_d^2}}
\label{eq:primpro}
\end{eqnarray}
The SSD of post-diffusion secondary electrons can be derived by converting the diffused 
proton distribution into a distribution of pre-diffusion electrons, and treating the latter as 
the source for diffusion of secondary electrons. Since electrons in the galactic disk lose 
essentially all (as is quantified in the next section) their energy before diffusing to the 
inner halo, it can be assumed that only the source electrons in the halo contribute 
to the population of the post-diffusion secondary electrons. Therefore, the spatial integrals 
over $\vec{r_m}$ can be performed in the $[-\infty,\infty]$ range, resulting in the expression
\begin{eqnarray}
f_e(\vec{r_o},t_o,\gamma)&=&\frac{1}{\pi^{3/2}}\frac{1}{A}\frac{16}{3}cnN_{0p}
\spp\times 0.14\frac{1}{P_e(\gamma)}\int_{t_m}^{t_o}dt_m(0.56
\frac{\gamma_t}{A})^{-\alpha_p}P(\gamma_t) \\ \nonumber
&&\int_0^{t_m}dt_i\frac{e^{\ell_p(t_m-t_i)(1-\alpha_p)}e^{-\nicefrac{x_o^2+y_o^2}{a^2+
4D(t_m-t_i)+r_d^2}-\nicefrac{z_o^2}{b^2+4D(t_m-t_i)+r_d^2}}}
{\left[a^2+4D(t_m-t_i)+r_d^2\right]\sqrt{b^2+4D(t_m-t_i)+r_d^2}},
\label{eq:secel}
\end{eqnarray}
where $A=(1/4)(m_{\pi}/m_e)\approx 70$ (Ramaty \& Lingenfelter 1966), $n$ is the 
gas number density, $N_{op}$, and $\alpha_p$ are the coefficient and power index of 
the proton injection spectrum, $\spp\approx35$ mb (obtained by averaging over an 
analytical approximation for the cross section of p-p interactions obtained for proton 
energies in the range 0.01-100GeV [Kelner \ea 2006]), is the mean cross section for p-p 
interactions, $\ell_p\equiv nc\kappa\spp$, and $\kappa\approx 0.45$, is the inelasticity 
of p-p interactions (Aharonian \& Atoyan 1996). 

Energy loss processes for energetic electrons are electronic excitations (in Coulomb 
scattering), bremsstrahlung, and Compton-synchrotron, with the total energy loss rate given by 
\be
P(\gamma)=p_{0} + p_{1}\gamma + p_{2} \gamma^{2}  \, ,
\label{eq:telr}
\ee
where $p_{0} \simeq 1.1\times 10^{-15}(n_{e}/10^{-3}cm^{-3}) \,s^{-1}$ is the loss 
rate by electronic excitations in fully ionized gas (Gould 1972); $n_{e}$ is the gas 
electron density. The coefficients of the bremsstrahlung and Compton-synchrotron rates 
(e.g., Blumenthal \& Gould 1970) are
$p_{1} \simeq 1.3\times 10^{-18}\times (n_{e}/10^{-3}cm^{-3}) \,s^{-1}$, and 
$p_2=s_0[(1+z)^{4}+0.1\times (B/(10^{-6}\mu G))^2] \,\,s^{-1}$, 
respectively, with $s_0$=1.3$\times 10^{-20}\, s^{-1}$ in the galactic halo, where 
Compton scattering is largely off the CMB (whose energy density increases steeply with 
redshift), unlike in the galactic disk where the stellar radiation field is stronger. The main 
energy loss of energetic protons is by interactions with gas protons at a rate $p(\gp)=
n_p(r)\,c\,\kappa\,\spp\gp$, where $\sinel$ is the mean cross section in inelastic scatterings.

\subsubsection{Disk Distributions}

In our approximate analytical calculations of \pe SSDs in the disk we use mean values of 
the gas density, magnetic field, and diffusion coefficient; doing so facilitates the computation 
of the radiative yields (e.g., emissivities). The primary electron SSD is then 
\begin{eqnarray}
f_e(t_o,\gamma)&=&\int_0^{t_o}\frac{1}{\pi^{3/2}}
\frac{\Delta \Dot{N}(\gt)P(\gt)}{P(\g)(a^2+r_d^2)\sqrt{b^2+r_d^2}}
\left[2\int_0^bdz_o\int_0^{a\sqrt{1-z_o^2/b^2}}2\pi x_odx_o\right]
e^{-\nicefrac{x_o^2}{a^2+r_d^2}-\nicefrac{z_o^2}{b^2+r_d^2}}\\
&=&\int_{\gamma_{min}}^{\gamma_{max}}\frac{d\gamma}{P(\gamma)}\int_{t_{min}}^{t_o}
\Delta \Dot{N}(\gt)P(\gt)\\
&\times&\left[\erf{\left(\frac{b}{\sqrt{b^2+r_d^2}}\right)-\frac{b}{\sqrt{b^2+r_d^2}}}
e^{-a^2/(a^2+r_d^2)\frac{}{}}\frac{\erf{(r_d\sqrt{1/(a^2+r_d^2)-1/(b^2+r_d^2)})}}{r_d\sqrt{1/(a^2+r_d^2)-1/(b^2+r_d^2)}}\right]
\end{eqnarray}
The corresponding expression for secondary electrons is
\begin{eqnarray}
f_e(t_o,\gamma)&=&\frac{1}{A}\frac{16}{3}cnN_{op}
\spp\times 0.13957\int_{\gamma}\frac{d\gamma}{P_e(\gamma)}
\int_{t_m}^{t_o}dt_m(0.558281\frac{\gamma_{t}}{A})^{-\alpha_p}P(\gamma_{t}) \\ \nonumber
&&\int_0^{t_m}dt_i\frac{e^{\ell_p(t_m-t_i)(1-\alpha_p)}}{\sqrt{b^2+4D(t_m-t_i)+r_{d}^2}}
\left[\sqrt{b^2+4D(t_m-t_i)+r_{d}^2}
\erf{\left(\frac{b}{\sqrt{b^2+4D(t_m-t_i)+r_{d}^2}}\right)} \right. \nonumber \\
&& \left. -be^{-a^2/[a^2+4D(t_m-t_i)+r_{d}^2]}
\frac{\erfi{\sqrt{a^2/[a^2+4D(t_m-t_i)+r_{d}^2]}}}{\sqrt{a^2/[a^2+4D(t_m-t_i)+r_{d}^2]}}\right],
\end{eqnarray}
where $\erfi$ is the imaginary error function. 

The galactic disk is modeled as an oblate ellipsoid with semi-major and semi-minor axes 
measuring 20\,kpc and~4\,kpc, respectively. Mean values of the magnetic field in the 
disks of NGC 4631 and NGC 4666 were observationally estimated to be 10 and 14 $\mu$G,
respectively (Stein 2017). The assumed disk mean gas density, $n=0.3$ cm$^{-3}$, 
corresponds to a total (disk) gas mass of approximately 3$\times$10$^{10}$ solar masses. 
Finally, the diffusion coefficient D is set to 3.4$\times$10$^{28}$ cm$^{2}$s$\,^{-1}$, in 
accord with the value taken by Strong \ea in their plain diffusion model 2. The radio 
luminosity generated by the disk at 1.575 GHz, as well as the hard X-ray (10 keV-200 keV), 
and $\gamma$ (100 MeV-100 GeV) luminosities, are calculated by integrating the respective emissivities over the spectral distribution functions of primary and secondary electrons, and 
neutral pions, with the latter derived from the corresponding proton distribution function, Eq.~(\ref{eq:grpd1}).

\subsubsection{Halo Distributions}

As we noted, our description of  energetic protons and electrons in the disk, which is 
similar to the more detailed work of Strong \ea (2010), provides the basis for the first 
quantitative modeling of the \pe distributions in the halo. We adopt the commonly 
used $\beta$-King model to describe the spatial profile of the gas density in the halo 
selecting values of the core radius and the index $\beta$ in the observationally deduced 
ranges, with the central density as a dependent parameter whose value is determined 
from the total gas mass. The magnetic field profile in the halo is related to that of the 
gas density by assuming conservation of magnetic flux  -- appropriate in a highly 
electrically conductive, fully ionized halo gas -- implying the scaling $\sim$n$^{2/3}$.

We divide the halo (excluding the galactic disk) into two separate regions: an inner 
cylindrical region with a radius of $20$ kpc and height that extends from r$_z$=4 kpc to 
20 kpc. The $\beta$-King model is specified by the halo parameters (Table~\ref{tab:prm}), 
with the z-dependence of the magnetic field and density profiles given by $\left[1+(r_z/r_c)^2\right]^{-3\beta/2}$. The outer halo region (at $r\geq 20$ kpc) is taken to be 
spherical with a density profile $\left[1+(r/r_c)^2\right]^{-3\beta/2}$.

Based on (rough) estimates of the mean strength of the IG field, we set a lower limit of 
$30$~nG on its value, which  is relevant only in the outer halo region. These analytic 
distributions allow the construction of 3-dimensional emissivity grids, thereby significantly 
simplifying integrations along the los and convolution with a Gaussian beam to produce 
radio brightness profiles and volume-integrated emissivities to compute total luminosities 
in the radio, hard X-ray, and $\gamma$ bands. The emissivity grid for the halo extends 
from the outer surface of the (model) disk out to $x_o=y_o=z_o=200$~kpc.

\subsection{Particle Radiative Yields}

The radiative yields of primary electrons, protons, and secondary electrons (and positrons) 
can be computed by integrating the relevant radiation emissivities over the particle distributions
detailed in the previous subsection. Accordingly, the spectral radio emissivity due to a population 
of relativistic electrons is (Blumenthal \& Gould 1970)
\begin{equation}
\frac{dE}{d\nu dt}=\frac{\sqrt{3}ke^3B}{4\pi m_e c^2}\int d\Omega_{\alpha}
N_{e}\sin{\alpha}\times\int_{\gamma_1}^{\gamma_2}f(\vec{r_o},t_o,\gamma)
d\gamma\frac{\nu}{\nu_c} \int_{\nu/\nu_c}^{\infty}K_{5/3}(\xi)d\xi, 
\label{eq:sync}
\end{equation}
where the electron density $N_{e}$ generally depends also on the pitch angle, $\alpha$, 
$\gamma_1$ and $\gamma_2$ bracket the electron spectral distribution, $\nu_c$ is the 
cyclotron frequency, $K_{5/3}$ is the modified Bessel function of $5/3$ order, and the 
physical constants have their usual meaning.   

The (photon) emissivity due to Compton scattering of low-energy CMB photons off 
relativistic electrons derived in the Thomson limit is (Blumenthal \& Gould 1970) 
\begin{equation}
\frac{dN_{\gamma,\epsilon}}{d\epsilon_1dt}=\frac{\pi r_0^2c}{2\gamma^4}
\frac{n(\epsilon)d\epsilon}{\epsilon^2}\left(2\epsilon_1
\log{\frac{\epsilon_1}{4\gamma^2\epsilon}}+\epsilon_1+4\gamma^2\epsilon
-\frac{\epsilon_1^2}{2\gamma^2\epsilon}\right),
\end{equation}
where $\epsilon$ and $\epsilon_1$ are the pre- and post-scattering photon 
energies, respectively. The electron classical radius is $r_0\equiv e^2/m_ec^2$ , 
and $n(\epsilon)$ is the spectral energy density of the scattering  radiation 
fields in the disk, consisting, in addition to the CMB, of diluted Planckian distributions 
at temperatures 3500K (starlight), and 30K (dust). 
Some 70\% of the Galactic stellar radiation field originates in a `thin' disk and the rest in 
a `thick' disk, with the former modeled as an ellipsoid with semi-major and semi-minor 
axes of 10 kpc and 300 pc (e.g., Gilmore \& Reid 1983), and the latter is the remaining disk 
volume.

By equating the stellar luminosity (in either the optical or IR bands) with the surface-integrated
flux for an isotropic diluted blackbody, $F=4\delta\sigma_bT^4$, where $\delta$ is the dilution 
factor and $\sigma_b$ is the Stefan-Boltzmann constant, the respective dilution factors can be computed for the two disk regions. The dilution factors for the stellar and dust fields are 
determined based on the estimate that of the total Galactic stellar luminosity, $\sim 5\times$10$^{10}$L$_{\odot}$ (Strong \ea 2010), 20\% is processed into the IR by 
interstellar dust. The Compton energy loss rate in the stellar radiation field is 
\begin{equation}
\frac{dE}{dt}=\frac{4}{3}\sigma_Tc\frac{4\delta\sigma_bT^4}{c}\gamma^2,
\end{equation}
where $\sigma_T$ is the Thomson cross section. Integrating over the energy $\epsilon$ and the scattering electron population yields the total Compton emissivity
\begin{equation}
\frac{dE_{tot}(\epsilon_1,\vec{r_o})}{d\epsilon_1dt}=
\int_{\gamma_1}^{\gamma_2}\int_{\epsilon_1/4\gamma^2}^{\infty}
\frac{dN_{\gamma,\epsilon}}{d\epsilon_1dt}\epsilon\times f(\vec{r_o},t_o,\gamma) d\epsilon d\gamma.
\end{equation}

The spectral $\gamma$-ray emission generated by $\pi^0$ decay can be calculated as 
(Mannheim \& Schlickeiser 1994):
\begin{equation}
f_{\gamma}(E_{\gamma},\vec{r_o})=2E_{\gamma}\int_{E_{\gamma}+\left[\left(m_{\pi^0}c^2)^2/
4E_{\gamma}\right)\right]}
^{\infty}\frac{f_{\pi^0}(E_{\pi^0},\vec{r_o})dE_{\pi^0}}{\sqrt{E_{\pi^0}^2-m_{\pi^0}^2c^4}},
\label{eq:grpd2}
\end{equation}
where (Pfrommer \& En\ss{}lin 2004)
\begin{equation}
f_{\pi^0}(E_{\pi^0},\vec{r_o})=8cnf_p(\vec{r_o},t_o,\gamma)\spp
\left(\frac{6E_{\pi^{\pm}}}{GeV}\right)^{-\frac{4}{3}(\alpha-\frac{1}{2})} GeV^{-1}.
\label{eq:grpd1}
\end{equation}

\subsection{Model Parameters}

As noted already, our approximate description of energetic \pe propagation, gas 
density and magnetic field distributions in the Galactic disk is chosen such that 
it is in general agreement with the more detailed GALPROP models of Strong \ea 
(2010). This is affected by selecting key parameter values that are reasonably close 
to those of the second diffusion model which is fully specified in the latter paper. 
We emphasize again that the main goal in our modeling of the disk is mostly to 
provide a realistic basis for determining \pe distributions and their yields in the 
Galactic halo where gas and magnetic field properties are poorly known, and can 
only be modeled by considering reasonable ranges of the key parameters. 

As can be seen in table~(\ref{tab:prm}), we consider 4 galactic halo models, each 
of which is a variation of the 4 basic parameters ($r_c$, $\beta$, n, D), where 
$M_g$ is the halo gas mass. We use a single model to characterize the disk, as 
discussed in the previous subsection.
\vskip 3em
  
\begin{table}
\center
\begin{tabular}{cc|c|c|c|c|c|}
\cline{3-7}
& & r$_c$ (kpc) & $\beta$ & n (cm$^{-3}$) & D (cm$^2$ s$^{-1}$) & M$_g$ (M$_{\odot}$) \\
\cline{1-7}
\multicolumn{1}{|c|}{\multirow{4}{*}{NGC 4631} }& model 1 & 2 
& .85 & 1 & 3$\times 10^{30}$ & 3$\times 10^{10}$   \\
\multicolumn{1}{ |c|}{\multirow{4}{*}{} }& model 2 & 3
& .85 & .4  & 3$\times 10^{30}$ & 3$\times 10^{10}$  \\
\multicolumn{1}{ |c|}{\multirow{4}{*}{} }& model 3 & 2
& .85 & 1 & 6$\times 10^{30}$ & 3$\times 10^{10}$  \\
\multicolumn{1}{ |c|}{\multirow{4}{*}{} }& model 4 & 2
& .7 & .25 & 3$\times 10^{30}$ & 3$\times 10^{10}$  \\
\hline
\multicolumn{1}{ |c|}{\multirow{4}{*}{NGC 4666} }& model 1 & 2
& 1.05 & 4 & 3$\times 10^{30}$ & 3$\times 10^{10}$  \\
\multicolumn{1}{ |c|}{\multirow{4}{*}{} }& model 2 & 3
& 1.05 & 1.3 & 3$\times 10^{30}$ & 3$\times 10^{10}$  \\
\multicolumn{1}{ |c|}{\multirow{4}{*}{} }& model 3 & 2
& 1.05 & 4 & 6$\times 10^{30}$ & 3$\times 10^{10}$  \\
\multicolumn{1}{ |c|}{\multirow{4}{*}{} }& model 4 & 2
& .85 & 1.1 & 3$\times 10^{30}$ & 3$\times 10^{10}$  \\
\hline
\end{tabular}
\caption{Model parameters}
\label{tab:prm}
\end{table}

\section{Results}

\begin{figure}
\centering
\epsfig{file=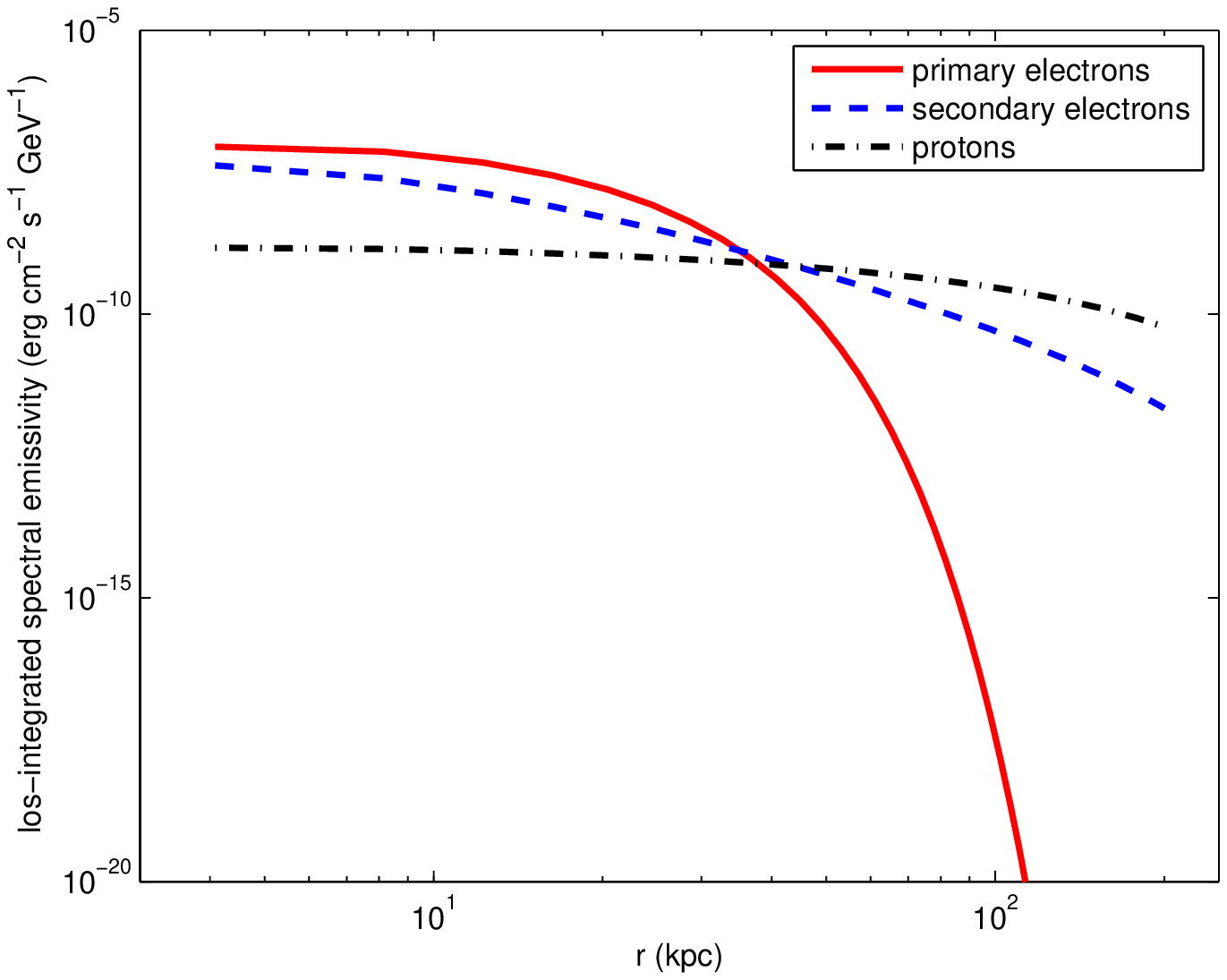,width=8.5cm,height=8.5cm,clip=}
\epsfig{file=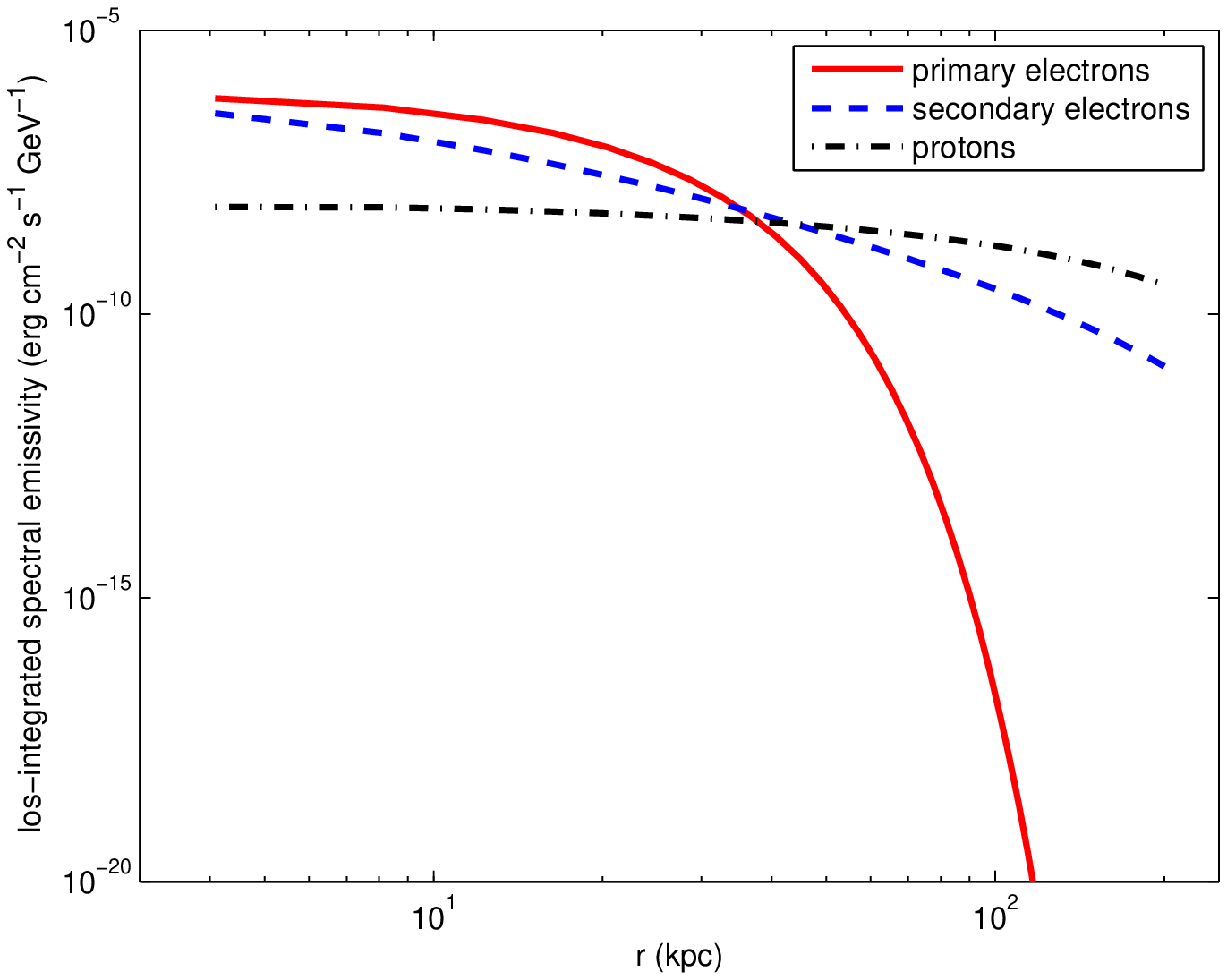,width=8.5cm,height=8.5cm,clip=}
\caption{
Time-averaged los-integrated particle surface emissivity profiles above the galactic 
plane, computed for $\gamma_e=10^4$ (primary and secondary electrons) and 
$\gamma_p\approx 12$ (protons). Results are shown for NGC 4631 (left) and NGC 
4666 (right)}
\label{fig:part_prof}
\end{figure}

\begin{figure}
\centering
\epsfig{file=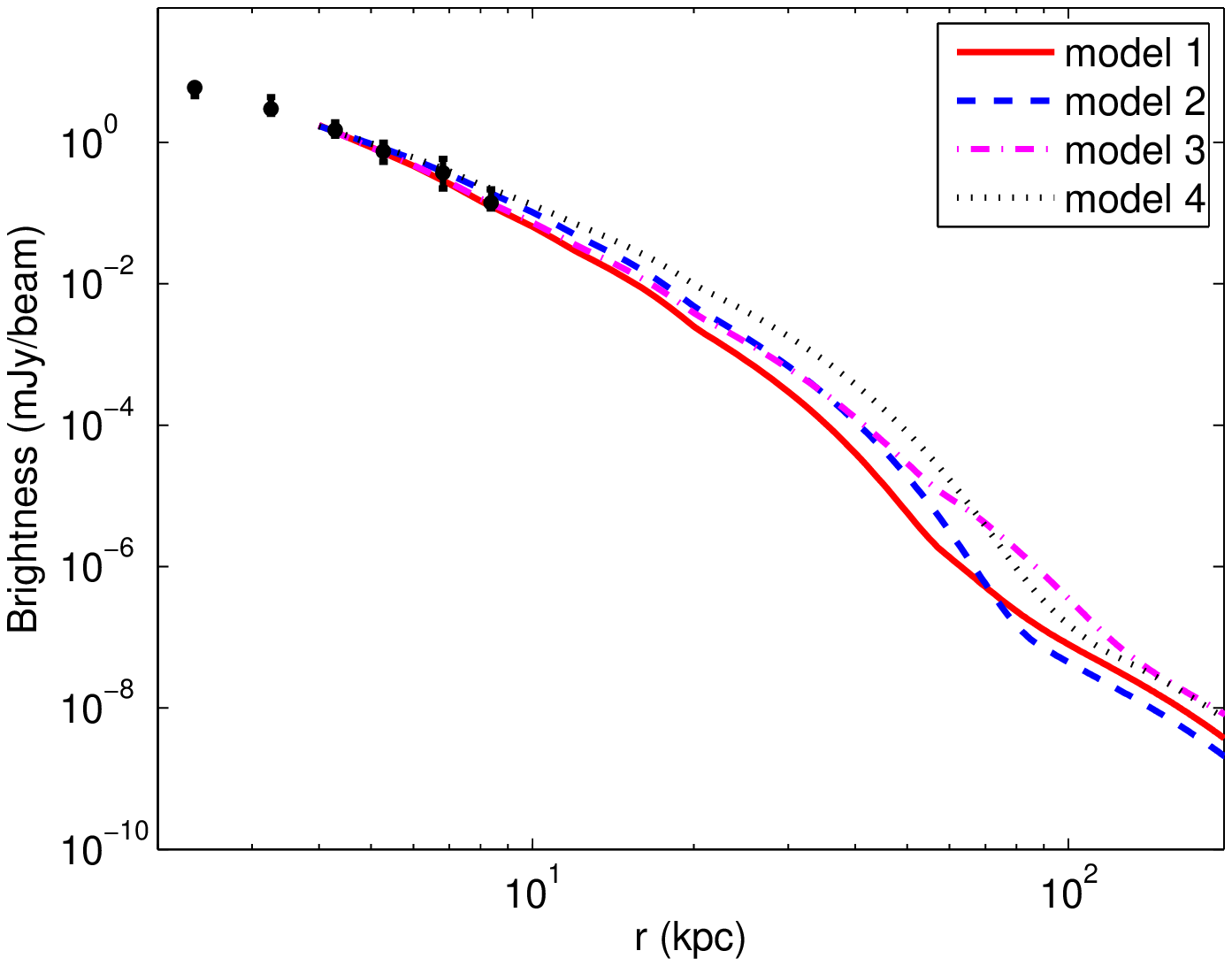,width=8.5cm,height=8.5cm,clip=}
\epsfig{file=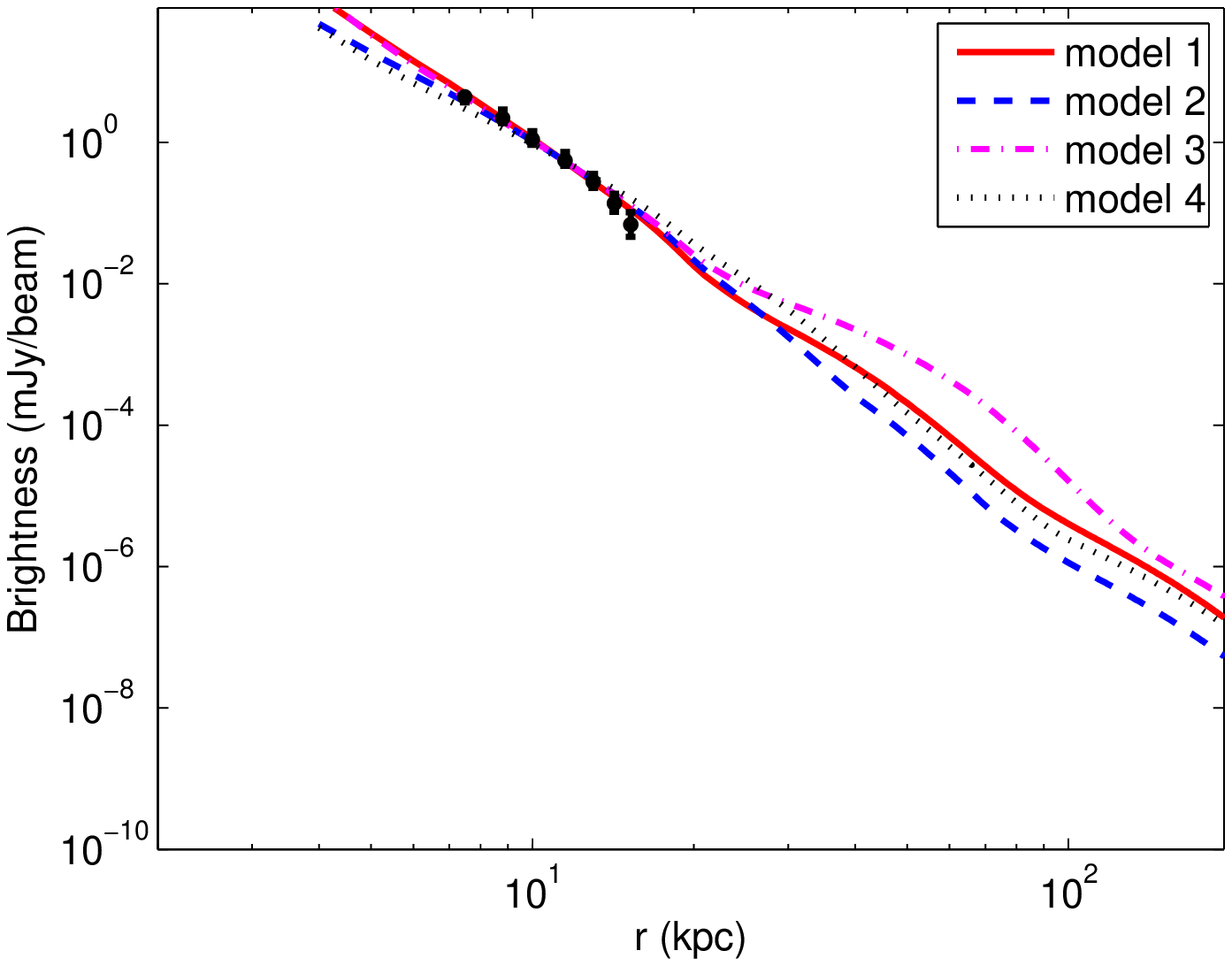,width=8.5cm,height=8.5cm,clip=}
\caption{Radio-brightness profiles at $\nu=1.575$ GHz, smoothed with a 
Gaussian beam of FWHM 18''. Results are shown for NGC 4631 (left) and 
NGC 4666 (right). Parameter values are listed in table~(\ref{tab:prm}).}
\label{fig:rad_prof}
\end{figure}

\begin{figure}
\centering
\epsfig{file=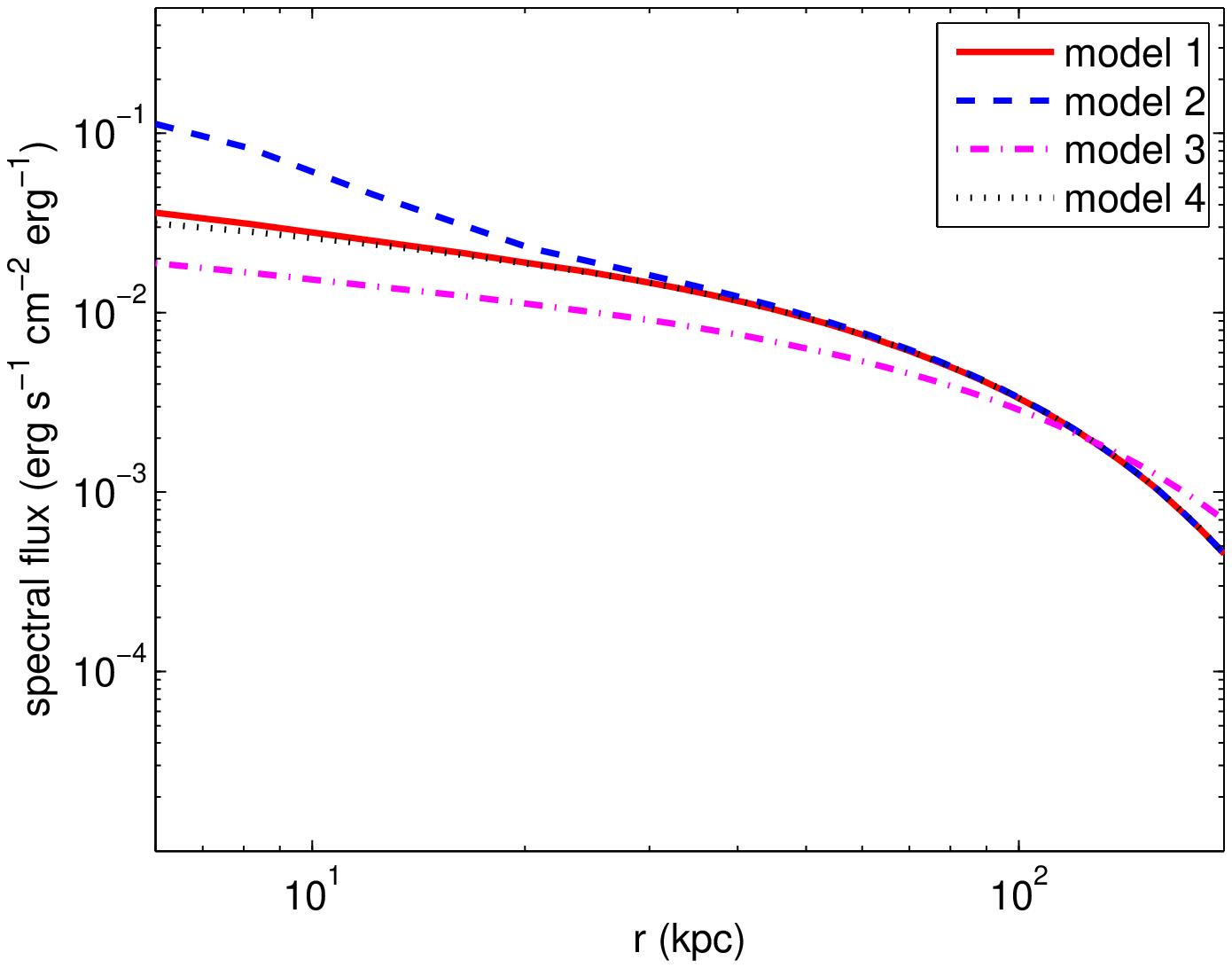,width=8.5cm,height=8.5cm,clip=}
\epsfig{file=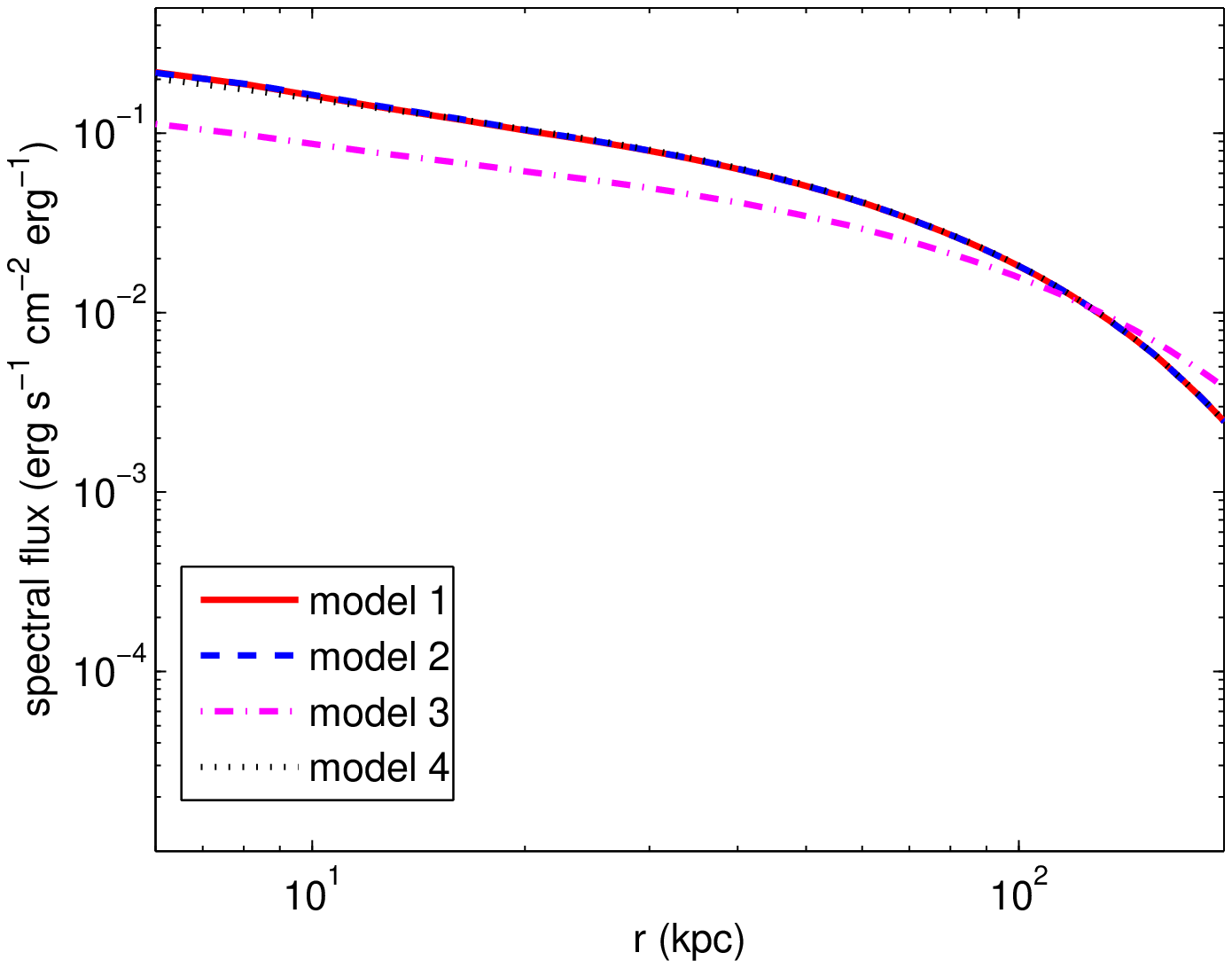,width=8.5cm,height=8.5cm,clip=}
\epsfig{file=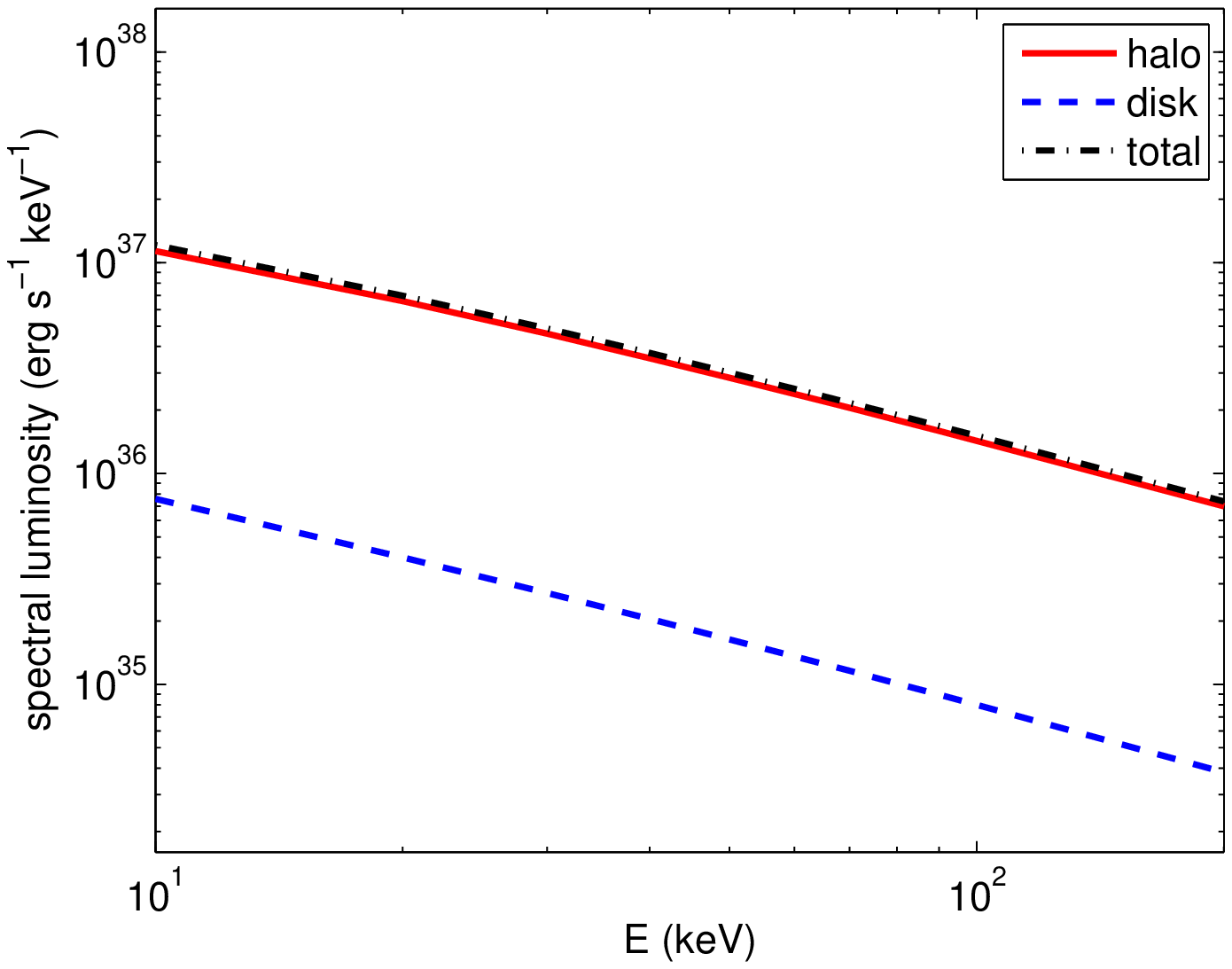,width=8.5cm,height=8.5cm,clip=}
\epsfig{file=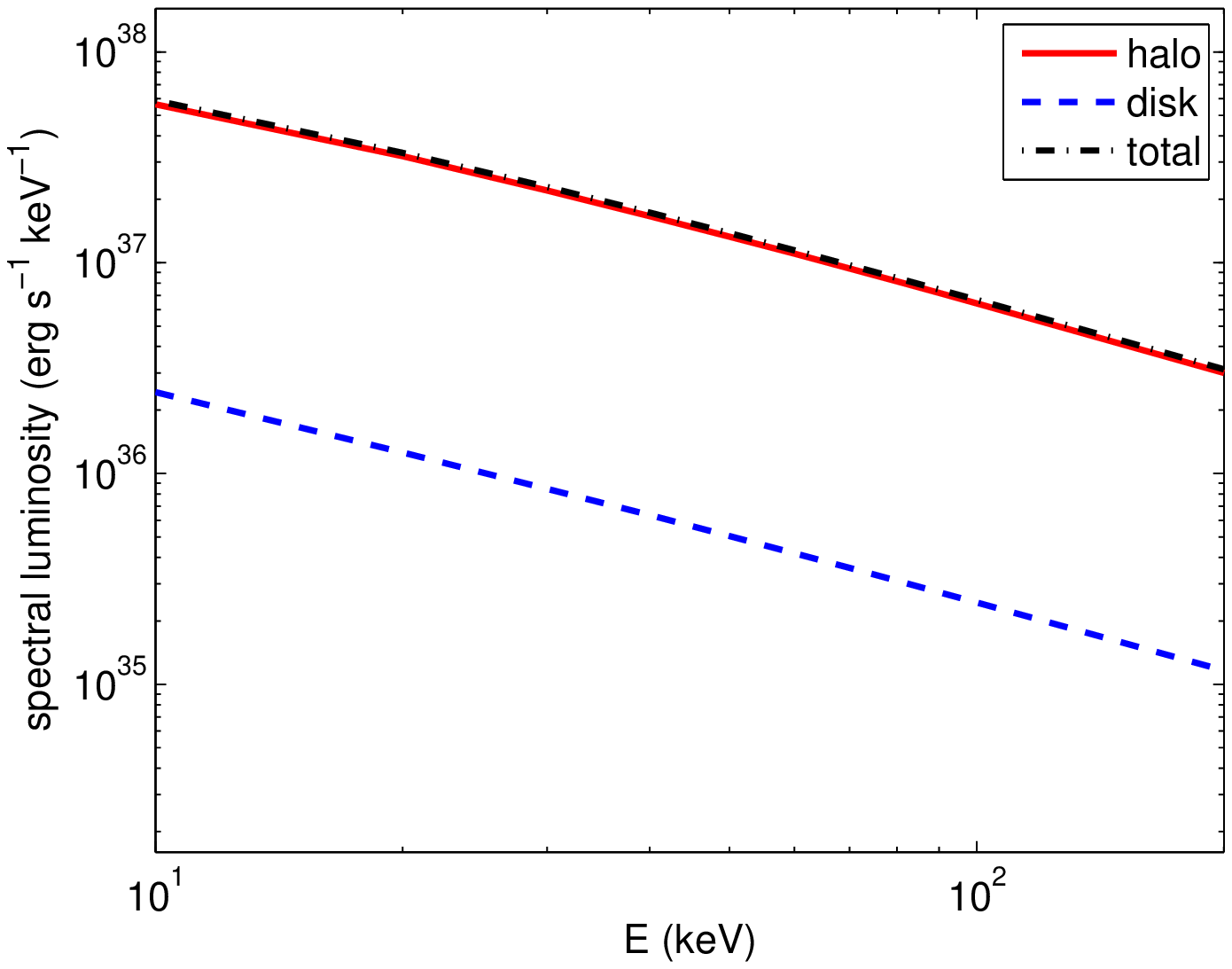,width=8.5cm,height=8.5cm,clip=}
\caption{X-ray emission from NGC 4631 (left panels) and NGC 4666 (right 
panels). Upper panels: los-integrated hard X-ray emissivity profiles at 20 keV. 
Lower panels: 10-200 keV spectral luminosities of the disk and halo.}
\label{fig:xray_prof}
\end{figure}

\begin{figure}
\centering
\epsfig{file=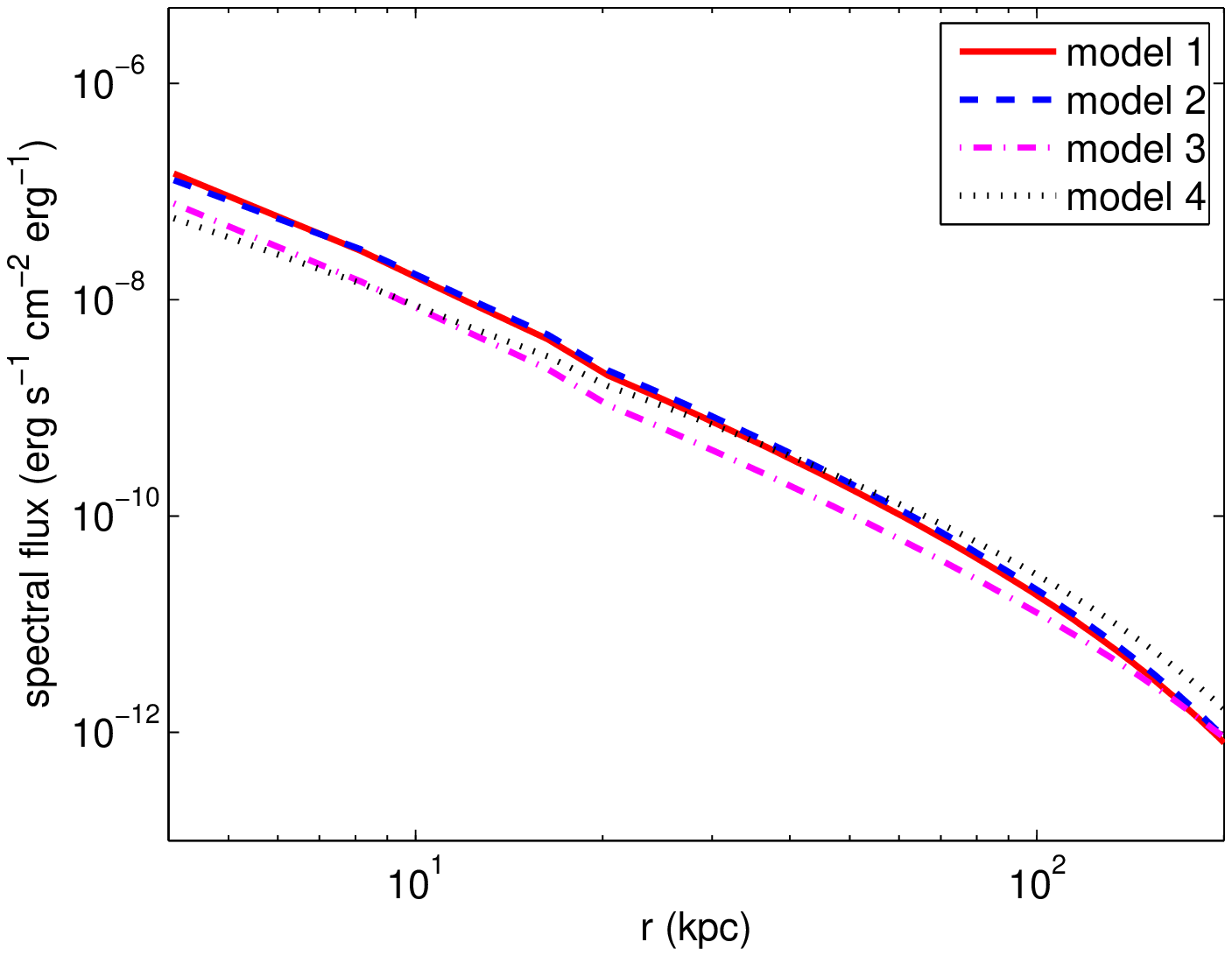,width=8.5cm,height=8.5cm,clip=}
\epsfig{file=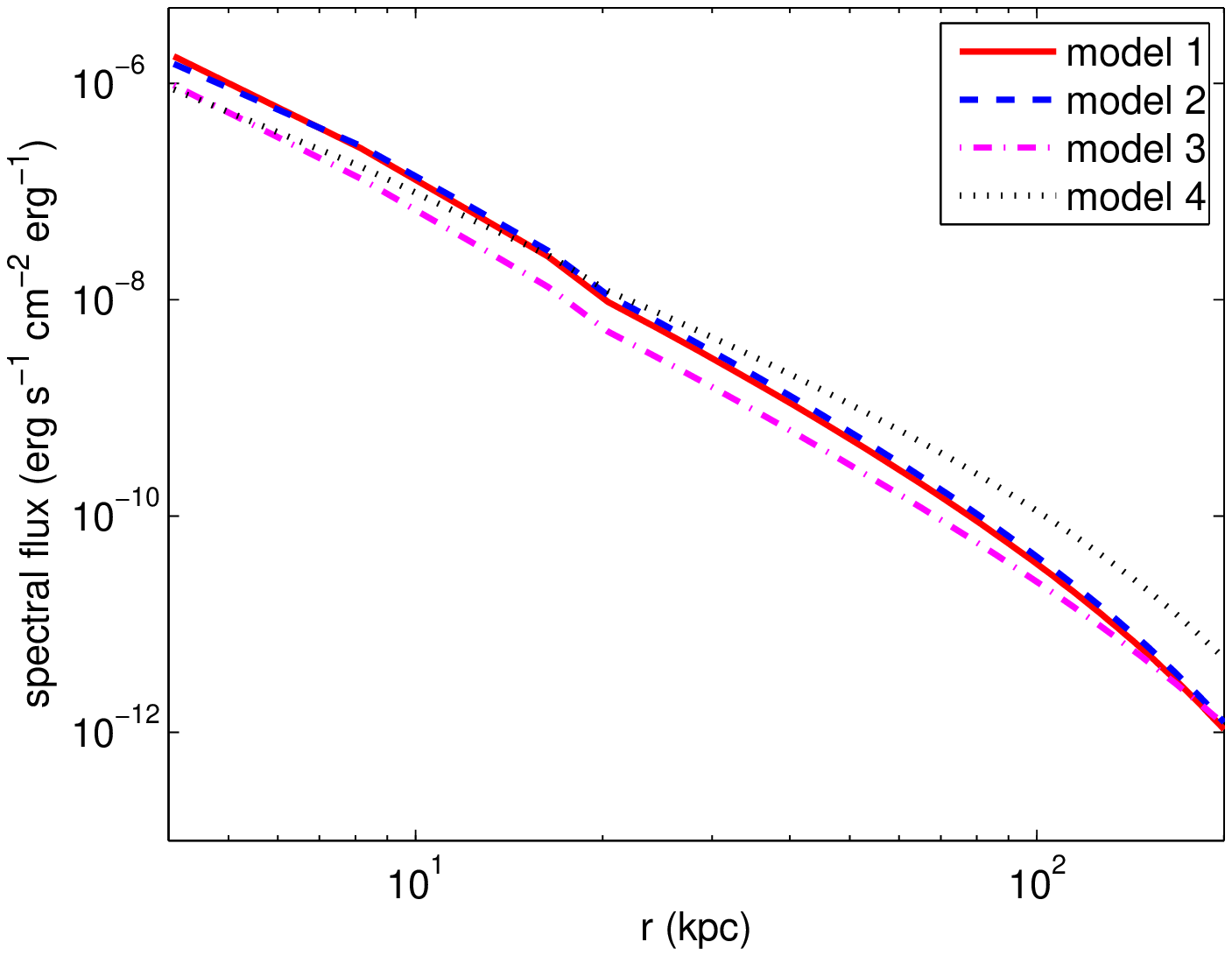,width=8.5cm,height=8.5cm,clip=}
\epsfig{file=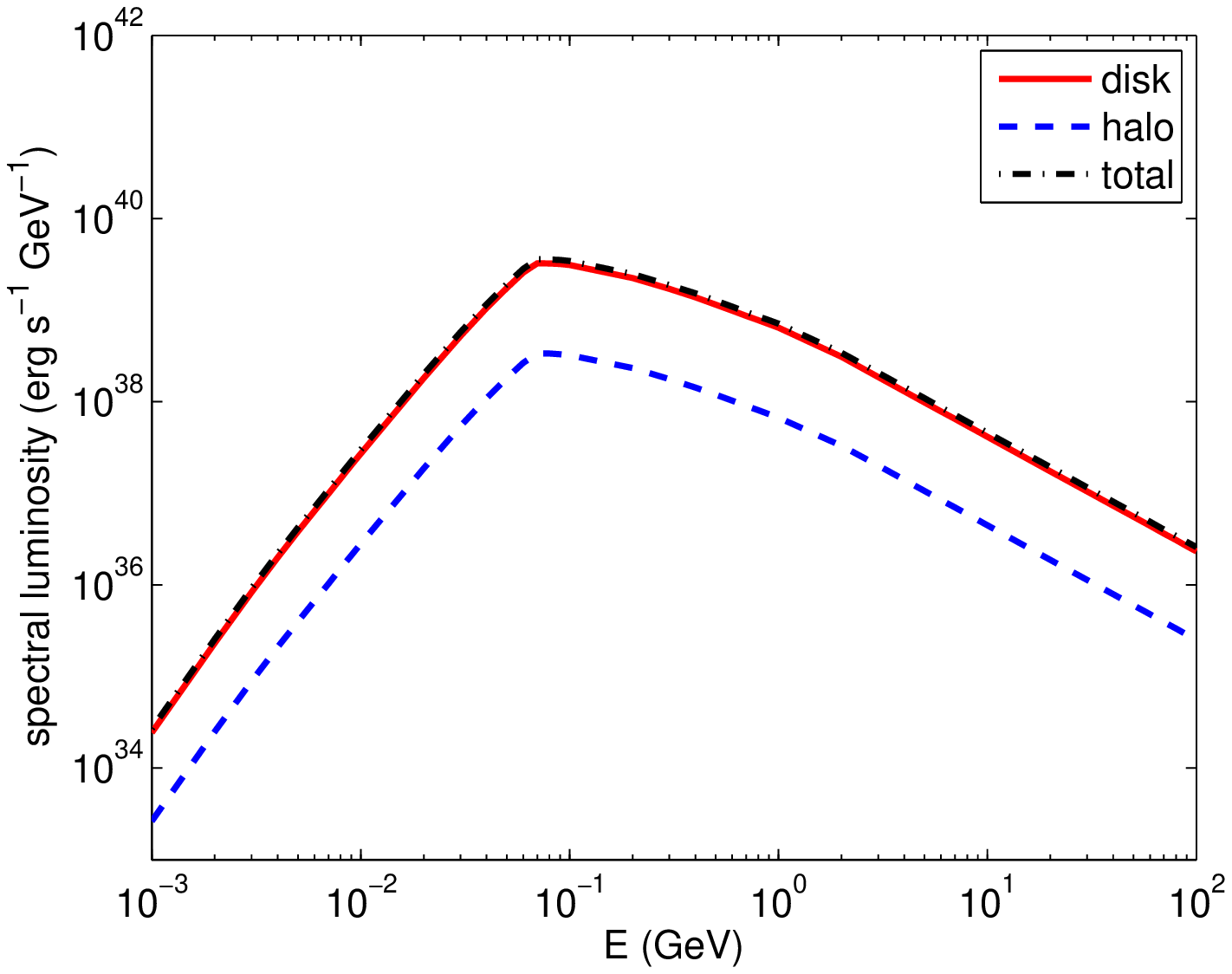,width=8.5cm,height=8.5cm,clip=}
\epsfig{file=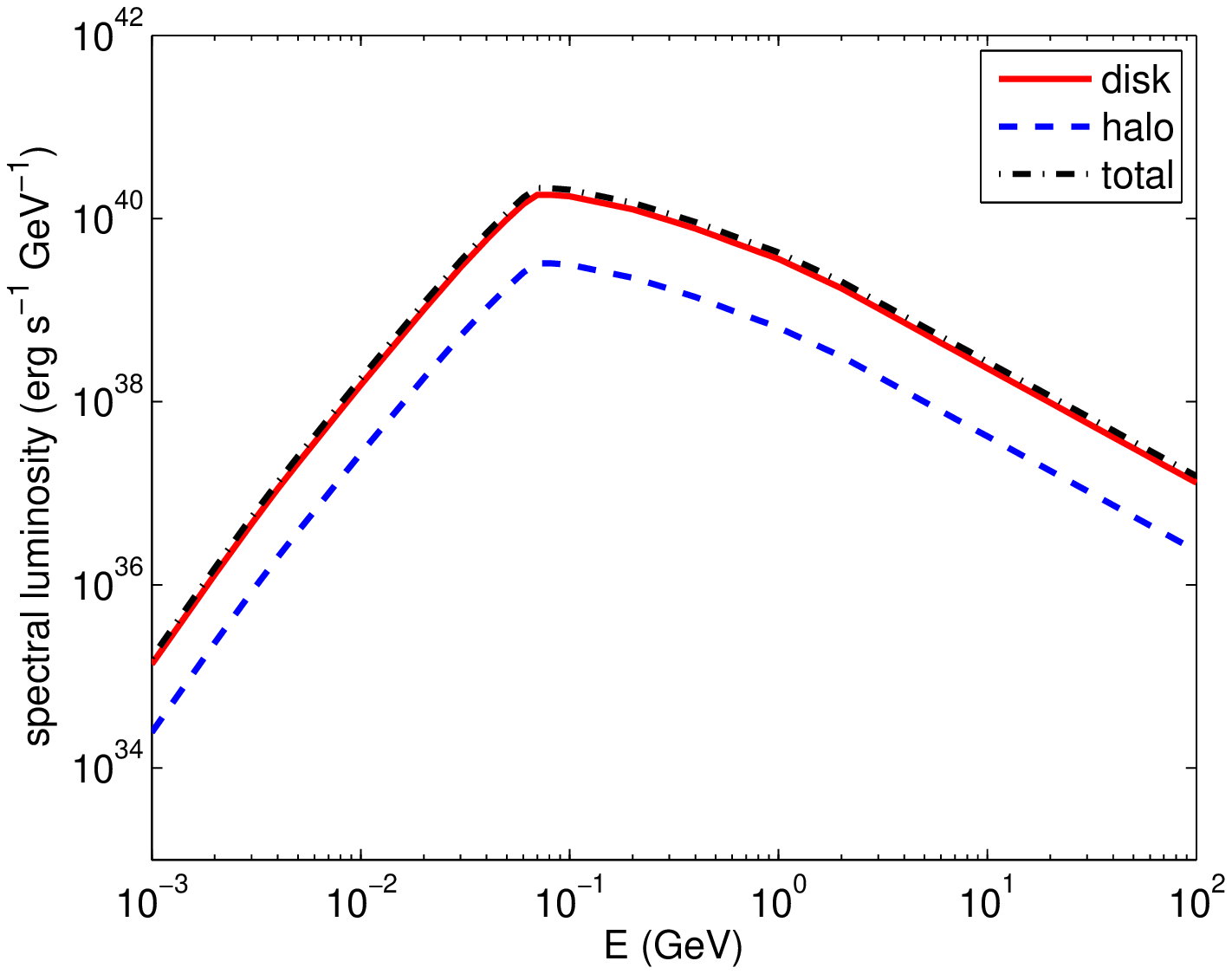,width=8.5cm,height=8.5cm,clip=}
\caption{ $\gamma$-ray emission from NGC 4631 (left panels) and NGC 4666 (right 
panels). Upper panels: profiles of the spectral flux at 100 MeV. Lower panels: 
1 MeV-100 GeV spectral luminosities in the disk and halo.}
\label{fig:gam_prof}
\end{figure}

We applied our modeling approach to predict the levels and spatial profiles of \pe 
radiative yields in the Galactic halo and in the halos of the two star forming, edge-on 
galaxies NGC 4631 and NGC 4666 at the respective distances of 7.4 and 27.5 Mpc, 
and with star formation rates (SFRs) of 1.33 and 7.29 M$_{\odot}$yr$^{-1}$ (Wiegert 
\ea 2015). The particle source distribution in the latter two galaxies was assumed to 
be as in the Galaxy, and the overall injection rates were determined by scaling up the 
Galactic rate, $\simeq 1$ M$_{\odot}$yr$^{-1}$, by the respective SFR of each 
galaxy. As shown in table~(\ref{tab:prm}), models 1-4 differ only in the value of a 
single parameter, with the central gas number density adjusted so as to yield the same 
halo gas mass of 3$\times 10^{10}$ M$_{\odot}$.

The spatial profiles of primary and secondary electrons at $\gamma_e=10^4$ (contributing 
most of the synchrotron emission at $\nu$=1.5 GHz), and protons at $\gamma_p\approx 12$ 
(kinetic energy of $\sim$10.3 GeV, above the threshold energy for pion production), are 
illustrated in Fig.~(\ref{fig:part_prof}); these are essentially Eqs.~(\ref{eq:primel}),
~(\ref{eq:primpro}), and~(\ref{eq:secel}), multiplied by the respective particle energy. 
The enhanced population of protons and secondary electrons above the galactic plane is 
clearly manifested, in clear contrast with the steeply falling profile of primary electrons.

For each galaxy we calculated the radio, hard X-ray, and $\gamma$ spectral luminosities 
and brightness profiles at 1.575 GHz, 20 keV, and 100 MeV, as well as spectra and 
bolometric luminosities in the 10-200 keV and 1 MeV-100 GeV bands. Radio brightness 
profiles (in mJy/beam) for the 4 models are plotted in Fig.~(\ref{fig:rad_prof}) for NGC 4631 
(left-) and NGC 4666 (right-hand panel). Intersections of the brightness contours measured 
by Wiegert \ea (2015) with a line perpendicular to the galactic plane and passing
through its centre were read off the brightness map; for each contour above and below the 
plane angular distances of the intersection points were converted into physical distances 
and averaged. Measurement errors were estimated as the square root of the contour 
rms noise and thickness added in quadratures. Model predictions, smoothed with a Gaussian 
beam of $\sim$18'' FWHM, were normalized either to the 3rd (NGC 4631) or 4th (NGC 4666) 
data point; these refer only to the halo, and consequently the two innermost measurements for 
NGC 4631, which still lie in the galactic disk, are irrelevant to the halo. The change of slope 
apparent in the brightness profiles at $\sim$30-80 kpc (depending on model) marks the height 
above (or below) the disk where the radio-yield of secondary electrons begins to dominate over 
that of primary electrons.

Keeping the same total halo gas mass with a higher core radius (model 2) and lower 
$\beta$ (model 4) implies lower central densities. This clearly results in reduced 
densities of secondary electrons, and consequently leads to the lower (with respect to 
model 1) radio brightness levels apparent in the secondary electron-dominated region of 
model 2; while the implied densities in model 4 are lower as well, the more moderate fall 
of the gas profile actually gives rise to a somewhat less steep brightness profiles, lying 
slightly above and below the model 1 profiles in the secondary-electron dominated zone 
for NGC 4631 and NGC 4666, respectively. By a qualitative comparison between the 
measurements and model predictions we infer that for NGC 4631 model 2 best matches 
the observations, whereas model 1 provides the closest fit for NGC 4666; these are referred 
to as ``fiducial'' models.

The los-integrated emissivity profile at 20 keV and the hard X-ray spectrum in the 10-200 
keV band (for the fiducial models) are illustrated in Fig.~(\ref{fig:xray_prof}). The profile 
essentially reflects that of the electrons since it primarily results from electron scattering off 
the (uniform) CMB. For NGC 4666 models 1, 3, and 4 are roughly similar, whereas for 
NGC 4631 the profile calculated for model 2 joins the model 1 and 4 profiles at $\sim$40 kpc.
The  emissivity profile for model 3 reflects the faster diffusion from the inner to the outer halo, 
which is clearly due to the higher diffusion coefficient in this model. As expected, the spectral 
X-ray luminosity in the disk and halo, shown in the lower panels of the figure, is essentially 
a power-law in energy, with spectral indices  in (the 10-100 keV band) of .93 and 1.00 
(NGC 4631), and .98 and 1.01 (NGC 4666) respectively. More generally, since particle 
acceleration is directly related to SF, the higher luminosities of NGC 4666 exactly reflect 
the higher (by a factor of 5.33) SFR in this galaxy than that in NGC 4631.

The los-integrated emissivity profile at 100 MeV and the $\gamma$-spectrum in the
1 MeV-100 GeV band are illustrated in Fig.~(\ref{fig:gam_prof}) for the fiducial models. 
Up to r$\sim$40 kpc (NGC 4631) and $\sim 20$ kpc (NGC 4666) above the galactic plane, 
the highest emission levels are predicted for models 1 (high central gas density) and 2 (large 
core radius); at larger distances the milder density slope of model 4 results in the highest 
emission in both galaxies. Also, the relatively low emission levels in model 3 reflect the 
high diffusion coefficient, which results in faster proton diffusion out of the inner higher 
density region of the halo. As can be clearly seen in the lower panels of the figure, the 
fraction of the $\gamma$ emission that originates in the halo is negligible with respect 
to that from the disk, essentially due to the lower gas densities outside the galactic disk.

The radio flux at 1.575GHz, hard X-ray luminosity in the 10-200 keV, and $\gamma$-ray
luminosity above 100 MeV are listed in table~(\ref{tab:res}), where the disk, halo, and total
contributions are explicitly specified. These results refer to the respective fiducial model of
each galaxy: model 2 for NGC 4631 and model 1 for NGC 4666. As is apparent from the table, 
the halo produces a non-negligible emission only in the hard X-ray band, constituting some 
93-95\% of the total luminosity of each galaxy.

Finally, we calculated the radiative efficiencies for particle energy conversion to the radio 
and $\gamma$-radiation channels, results of which are listed in table~(\ref{tab:calf}), and 
compared with the respective factors specified by Strong \ea (2010).

\begin{table}
\center
\begin{tabular}{cc|c|c|c|}
\cline{3-5}
& & hard X-ray (10-200 keV) & $\gamma$ ($\ge$ 100 MeV)& 
1.575 GHz flux \\
\cline{1-5}
\multicolumn{1}{ |c|}{\multirow{3}{*}{NGC 4631} }& disk & 2.5$\times$10$^{37}$ 
& 3.2 $\times$10$^{39}$ & 695 mJy \\
\multicolumn{1}{ |c|}{\multirow{3}{*}{} }& halo &  4.2$\times$10$^{38}$ 
& 3.4 $\times$10$^{38}$ & 36 mJy \\
\multicolumn{1}{ |c|}{\multirow{3}{*}{} }& total & 4.5$\times$10$^{38}$ 
& 3.5 $\times$10$^{39}$ & 731 mJy \\
\hline
\multicolumn{1}{ |c|}{\multirow{3}{*}{NGC 4666} }& disk & 7.8$\times$10$^{37}$ 
& 1.8 $\times$10$^{40}$ & 306 mJy \\
\multicolumn{1}{ |c|}{\multirow{3}{*}{} }& halo & 2.0$\times$10$^{39}$ 
& 3.3 $\times$10$^{39}$ & 3 mJy \\
\multicolumn{1}{ |c|}{\multirow{3}{*}{} }& total & 2.1$\times$10$^{39}$ 
& 2.1 $\times$10$^{40}$ & 309 mJy \\
\hline
\end{tabular}
\caption{Computed radiative yields of NGC 4631 and NGC 4666.}
\label{tab:res}
\end{table}

\begin{table}
\begin{tabular}{l|c|c|c|c|c|}
\cline{2-6}
 & \multirow{3}{*}{\makecell{Radio \\ (.001-100 GHz)}} & \multicolumn{2}{c|}{\multirow{2}{*}
{$\gamma$: Compton}} 
 & \multicolumn{2}{c|}{\multirow{2}{*}
{$\gamma$: $\pi^{0}$-decay}} \\
 &  & \multicolumn{2}{c|}{} & \multicolumn{2}{c|}{} \\ \cline{3-6} 
 &  & .01-100 MeV & \multicolumn{1}{c|}{100 MeV-100 GeV} & .01-100 MeV & 100 MeV-100 GeV \\ \hline
\multicolumn{1}{|c|}{NGC 4631} & .08 & .20 & .17 & .002 & .04 \\ \hline
\multicolumn{1}{|l|}{NGC 4666} & .09 & .20 & .15 & .002 & .04 \\ \hline
\multicolumn{1}{|l|}{Milky Way} & .16 & .14 & .14 & .0002 & .007 \\ \hline
\end{tabular}
\caption{Radiative efficiencies for NGC 4631 and NGC 4666, compared with those 
determined by Strong \ea (2010) for the Milky Way.}
\label{tab:calf}
\end{table}

\section{Discussion}

The main objective of our work has been a quantitative description of energetic 
\pes in halos of SFGs, based on detailed modeling of their SSDs in galactic disks. We 
adopted an approximate diffusion solution to particle distributions in the halo, 
normalized at the disk boundary by using detailed results from GALPROP models for the 
Galaxy.  Our treatment provides a realistic basis for estimating the particle radiative yields 
in the halo for an assumed magnetic field profile, halo gas mass, gas core radius, and 
diffusion coefficient. Doing so we bracket the most likely ranges of spectral and spatial 
profiles of halo radio, hard X-ray, and $\gamma$-ray emission; these constitute a tangible 
basis for assessing the detection feasibility of the integrated fluxes in these spectral bands. 

Results of our calculations indicate that emission from the halo of a SFG will be 
comparable to or higher than that of the disk only in the hard X-ray band. 
This clearly is due to the relatively high fraction of secondary electrons in the halo 
and the uniformity of the CMB (the dominant radiation field), in sharp contrast to the 
steeply decreasing magnetic field and gas density that determine emission levels 
in the radio and $\gamma$-ray, respectively. 
In our `fiducial' models the estimated 10-200 keV halo luminosities of NGC 4631 
and NGC 4666 are $\sim 17$ and $\sim 26$ times higher than the respective (NT) disk 
luminosities. For each of these galaxies the predicted luminosities in the other three 
models are within a factor of two of the respective values specified in Table 2. 
While high, the total 10-200 keV luminosity of even the (more luminous) SB 
galaxy NGC 4666, $\simeq 2\times 10^{39}$ erg/s, is still substantially lower than 
stellar X-ray emission from the disk, which is dominated by high-mass X-ray binaries 
(but includes also significant emission from low-mass X-ray binaries and thermal 
emission from diffuse IS gas). The disk emission can be estimated using the 
observationally deduced scaling relation (Mineo \ea 2012) for the 0.5-8 keV 
luminosity in terms of the SFR, which for NGC 4666, yields a value that is more than 
a factor of 10 higher than the predicted 10-200 keV luminosity of the halo. 
We note that for the Galaxy, with a global SFR of $\sim 1$  M$_{\odot}$yr$^{-1}$, 
Strong \ea (2010) estimates for the (total) disk luminosities are $L(10 keV - 100 MeV) 
\simeq 2.2\times 10^{38}$ erg/s, and $L(0.1 - 100 GeV) \simeq 7.2\times 10^{38}$ erg/s 
for their diffusion model 2.

Our estimates of hard X-ray luminosities of the two galaxies provide sufficient 
motivation for dedicated searches for halo emission by the next generation hard 
X-ray satellites which will likely have sufficient sensitivity and angular resolution 
to separate out disk and halo emission in nearby edge-on galaxies. Prime candidates 
for such observations are the two nearby, nearly edge-on SBGs NGC253 and M82, 
which at 2.5 and 3.6 Mpc, respectively, are more than twice closer than NGC 4631, 
and with significantly higher SF rates, are more optimal targets for mapping X-ray 
emission from their halos. The Fermi-LAT telescope detected emission above 200 
MeV from NGC253 and M82 (Abdo \ea 2010); the emission was determined to 
be non-varying and point-like. 
A recent (`Pass 8') analysis of 7-year Fermi-LAT data on a sample of SFGs (Rojas-Bravo 
and Araya 2016) resulted only in upper limits on the emission above 100 MeV. The sample 
includes NGC 4631 for which a 95\% upper limit of $5\times 10^{39}$ erg/s was set. (This 
upper limit is a factor $\sim 1.8$ lower than the value deduced by Ackermann \ea 2012 in 
their analysis of emission from SFGs in the 3-year Fermi-LAT database.) The current 
bound is somewhat higher than our predicted total $\gamma$-ray luminosity. 
Given the intense SF activity in the central disk regions in these galaxies, the measurement 
of high energy X-and-$\gamma$-ray emission from these closest SBGs galaxies is not 
surprising, and indeed was predicted based on detailed models (e.g., Goldshmidt \& 
Rephaeli 1995, Domingo-Santamar\'{i}a \& Torres 2005; Persic, Rephaeli, \& Arieli 2008; 
de Cea del Pozo et al. 2009; Rephaeli, Arieli, \& Persic 2010). 

In assessing realistic ranges of key parameters in the estimation of the predicted radiative 
yields of star forming galaxies, we note that in a recent analysis (Heesen \ea 2018) of radio 
emission profiles in nearby galaxies, higher values of the SFRs and mean disk magnetic 
fields were determined. The latter analysis is based on a different approach to modeling 
energetic electron and magnetic field distributions in the thin and thick disk regions; key 
differences between the respective approaches include particle 
source distribution (Gaussian in our treatment, uniform across the galactic disk but 
vanishing elsewhere in that of Heesen \ea), the duration of the transport process (time 
integration is performed in our approach, whereas steady state is presumed in theirs), 
and our inclusion of radiative emission from secondary electrons which they (apparently) 
have not done. Nonetheless it is still relevant to assess the net effect of the higher values 
of these quantities on our results. For the two galaxies considered here, NGC 4631 and 
NGC 4666, SFRs of 2.9 and 16.2 M$_{\odot}$yr$^{-1}$ were used; these are about 
a factor $\sim 2.2$ higher than the values (from Wiegert \ea 2015) adopted here. The 
SFR constitutes an overall normalization of particle densities in the disk, but since we 
normalize the predicted radio (surface) emissivity to the observed profiles, a different 
SFR would only affect this overall normalization factor. Mean magnetic field values 
in the disks of the two galaxies were taken to be 13.5 and 18.2 $\mu$G, higher by 
about $\sim 1.3$ than our values. With the level of the overall B-dependence of the 
synchrotron emissivity, this difference amounts to about the same (latter) factor. 
Therefore, given inherent observational uncertainties, and the fact that our calculated 
radiative yields (for the set of models considered here) span appreciably wider ranges 
than that implied by the above SFR and magnetic field values, the overall impact on 
our predictions is small.

The ratio of the particle energy lost by interactions in IS space to the energy injected 
out of the acceleration sites is referred to as the calorimetric efficiency. This fraction is 
known to be very large ($\sim 90\%$) for primary electrons, but quite low ($\sim 10\%$) 
for protons. Obviously, additional losses in the halo result in essentially negligibly small 
residual primary electron component in the outer halo. Of interest is mainly the total 
calorimetric efficiency of protons and their residual energy content in the outer halo. 
For the models explored here, our estimates indicate that the total calorimetric efficiencies 
of protons are in the range $\sim .1-4\%$. Propagation through the halo increases 
the total proton energy losses by only a few percent over the losses in the disk.

An important consequence of the low calorimetric efficiency of protons diffusing out 
of SFGs is their accumulated density in clusters. In the central regions of rich clusters 
the gas density and magnetic fields are significantly enhanced with respect to their levels 
in IG space outside clusters, with typical values of O(10$^{-3}$) cm$^{-3}$ and O(1) 
$\mu$G, higher than corresponding values in the outer halo of a SFG that is not a cluster 
member. The higher concentration of SFGs in a cluster and stronger coupling of protons 
to the magnetized IC gas result in the creation of secondary electrons and the emission 
of significant radio and NT X-\&-$\gamma$ emission in the central cluster region.

We have recently carried out a first detailed assessment of galactic \pe in IC space (Rephaeli 
and Sadeh 2016) based on a calculation of the particle spectro-spatial distributions assuming 
that electrons diffuse out of radio and SF galaxies. Given the lack of unequivocal 
observational evidence for appreciable proton component in the lobes of radio galaxies, 
we conservatively ignored the contribution of radio galaxies assuming that protons 
originate only in SFGs. The extended distribution of SFGs, whose relative fraction 
increases with distance from the cluster center, is a key feature in accounting for the 
large size of radio halos. Particle escape rates were approximated by scaling to the 
estimated rates from the Galactic disk (Strong \ea 2010), with no account taken of 
propagation and energy losses in the halo. This approach was applied to conditions 
in the Coma cluster, where the number of SFGs was estimated from the total blue 
luminosity of the cluster, and only the two central powerful radio galaxies were 
included as electron sources. 

We found that for reasonable models of the gas density and magnetic field spatial 
profiles, the predicted profile of the combined radio emission from primary 
and secondary electrons is roughly consistent with that deduced from current 
measurements of the Coma halo. However, in this rather conservative approach (in 
accounting for \pe sources in clusters) the level of radio emission was found to be 
appreciably lower than the measured emission; this suggests that there could be 
additional particle sources, such as AGN and lower luminosity (than that of the 
two strong central) radio galaxies. Also calculated were the levels of NT X-ray 
emission, which was predicted to be mostly by Compton scattering of electrons 
from radio galaxies off the CMB, and $\gamma$-ray emission from the decay of 
neutral pions produced in interactions of protons from SFGs with protons in IC gas. 
Since the predicted levels of NT X-ray and $\gamma$-ray emission from (e.g.) the 
Coma cluster are appreciably lower than current upper limits, only weak constraints 
can at present be obtained on the steady state \pe IC distributions.

In assessing the likelihood of additional proton component in clusters, we note  
that the low calorimetric efficiency of protons, which are clearly accelerated in the 
nuclei of radio galaxies, implies that these could be important proton sources. 
Some evidence for significant proton populations in the lobes of radio galaxies comes 
from Fermi-LAT measurements of emission above 100 MeV from the nearby 
Centaurus \rm A (Abdo \ea 2010) and Fornax \rm A (Ackermann \ea 2016). 
Based on detailed models of the radio to $\gamma$ spectral energy distributions of these 
galaxies it was concluded that the measured $\gamma$ emission is from the $\pi^{0}$ 
decay (produced in p-p interactions), rather than Compton up-scattering of the 
radio-emitting electrons off the background optical radiation field (McKinley \ea 2015, 
Ackermann \ea 2016). However, at least in the case of Fornax \rm A, it was recently 
shown (Persic \& Rephaeli 2019) that the optical radiation field in the lobes (which is 
dominated by the central host galaxy NGC 1316) is sufficiently intense to account for 
the Fermi-LAT measurements by Compton scattering. It remains to be seen if the 
measured emission from the Coma halo can be fully accounted for by \pes from all 
SF and radio galaxies in the cluster. If this turns out to be unrealistic, then there 
would be stronger motivation to consider particle re-acceleration (in IC space) as an 
alternative model for maintaining the requisite level of radio-emitting electrons.

\section*{Acknowledgements}

This work has been supported by a grant from JCF (San Diego, CA). The authors wish to
thank Theresa Wiegert for clarifying details of the radio-map in the CHANG-ES 2015 paper, 
and the referee for a constructive report. 

\section{References}
\def\ref{\par\noindent\hangindent 20pt}

\ref Abdo A.A. et al., 2010, \apj, 719, 1433
\ref Ackermann M., et al., 2012, ApJ. 755, 164
\ref Ackermann M., et al., 2016, ApJ. 826, 1 
\ref Aharonian F.A., Atoyan A.M., 1996, \Aa, 309, 917
\ref Atoyan A.M., Aharonian F.A., \& V\"{o}lk H.J., 1995, \prd, 52, 6
\ref Blumenthal G.R., Gould R.J., 1970, \rmph, 42, 2
\ref de Cea del Pozo E., Torres D.F., Rodriguez Marrero A.Y., 2009, ApJ, 698, 1054
\ref Domingo-Santamar\'{i}a E., Torres D.F., 2005, A\&A, 444, 403
\ref Gilmore G., Reid N., 1983, \mn, 202, 1025
\ref Goldshmidt O., Rephaeli Y., 1995, \apj, 444, 113
\ref Gould R.J., 1972, Physica, 60, 145
\ref Heesen V. et al., 2018, \mn, 476, 158
\ref Kelner S.R., Aharonian F.A., Bugayov V.V., 2006, \prd, 74, 034018
\ref Mannheim K., Schlickeiser R., 1994, A\&A, 286, 983
\ref McKinley B., et al., 2015, \mn, 446, 3478
\ref Mineo S., Gilfanov M., Sunyaev R., 2012, MNRAS, 419, 2095 
\ref Orlando E., 2019, \prd, 99, 043007
\ref Orlando E., Strong A., 2013, MNRAS, 436, 2127
\ref Persic M. Rephaeli Y., Arieli Y., 2008, A\&A, 486, 143
\ref Pfrommer C., En\ss{}lin T.A., 2004, \Aa, 426, 777
\ref Ramaty R., Lingenfelter R.E., 1966, JGR, 71, 3687
\ref Rephaeli Y., Arieli Y., Persic M., 2010, \mn, 401, 473
\ref Rephaeli Y., Sadeh S., 2016, \prd, 93, 10
\ref Rojas-Bravo, C., Araya, M., 2016, MNRAS, 463, 1068 
\ref Stein Y., 2017, Magnetic Fields and Cosmic Ray Transport in Edge-On Spiral Galaxies \\ 
from the CHANG-ES Sample, Ruh-Universit\"{a}t Bochum
\ref Strong A.W., Porter T.A., Digel S.W., J\'{o}hannesson G., Martin P., Moskalenko I.V., 
Murphy E.J., Orlando E., 2010, \apjl, L58
\ref Syrovatskii S.I., 1959, Soviet Astronomy, 3, 22
\ref Wiegert T. et al., 2015, Astron. J. 150, 81

\bsp

\label{lastpage}

\end{document}